1

PACS:
52.55.Dy
52.65.Ff
94.30.Di# Technique of Examination of the Current Sheath Stability in the Electrically Neutral Approximation

V.V.Lyahov, V.M.Neshchadim

Institute of Ionosphere, Kamenskoe Plato, 050020 Almaty, Kazakhstan**Introduction**

Thus, despite success in the study of a large number of instabilities of magnetoactive plasma, development of research techniques to study stability of sharply inhomogeneous plasma structures, which should be regarded as self-consistent and be of arbitrary thickness (the characteristic thickness is determined by plasma and magnetic field parameters) is essential. Prior to attack on this physical problem we need to create necessary instruments: 1) to solve kinetic equation with a self-consistent electromagnetic field for perturbation of distribution function; 2) to calculate on the basis of this solution tensor of dielectric permeability of plasma of the current layer necessary for a material equation; 3) to derive a dispersion equation from the Maxwell system of equations closed with the obtained material equation, and develop a technique to analyze it

**Formulation of the Problem**

Let plasma concentrate in the plane z=0 (refer to Fig. 1) and antiparallel magnetic fields - in the upper and lower half-spaces. In this plane the self-consistent current sheath parting fields with antidirectional magnetic fields develops.

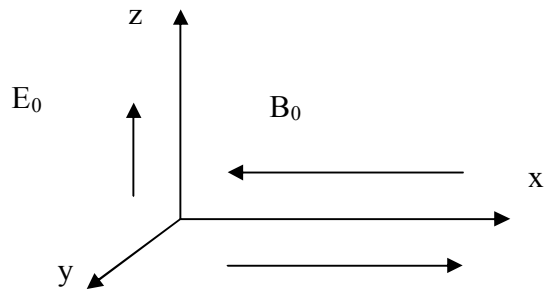

Fig. 1

The problem is one-dimensional and all quantities depend on variable z. The investigated medium is described by the system from the kinetic equation:

$$\frac{\partial f_\alpha}{\partial t} + \vec{v}\frac{\partial f_\alpha}{\partial \vec{r}} + e_\alpha \left\{ \vec{E} + [\vec{v}\vec{B}] \right\} \frac{\partial f_\alpha}{\partial \vec{P}_\alpha} = 0, \tag{1}$$

and Maxwell equations with a self-consistent electromagnetic field (no external sources)

$$rot\vec{B} = \frac{1}{c^2}\frac{\partial \vec{E}}{\partial t} + \frac{1}{\varepsilon_0 c^2}\vec{j}, div\vec{B} = 0, \quad (2)$$

$$rot\vec{E} = -\frac{\partial \vec{B}}{\partial t}, div\vec{E} = \frac{\rho}{\varepsilon_0},$$

Where:

$$\rho = \sum_\alpha e_\alpha \int \delta f_\alpha d\vec{P},$$

$$\vec{j} = \sum_\alpha e_\alpha \int \vec{v}\delta f_\alpha d\vec{P}.$$

The problems (1), (2) are solved by methods of perturbation theory:

$$f_\alpha(\vec{P},\vec{r},z,t) = f_{0\alpha}(\vec{P}) + \delta f_\alpha(\vec{P},\vec{r},z,t);$$
$$\vec{E}(\vec{r},z,t) = \vec{E}_0(z) + \delta\vec{E}(\vec{r},z,t); \quad (3)$$
$$\vec{B}(\vec{r},z,t) = \vec{B}_0(z) + \delta\vec{B}(\vec{r},z,t).$$

Plasma is considered to be weakly non-equilibrium $\delta f_\alpha(\vec{P},\vec{r},z,t) < f_{0\alpha}(\vec{P})$.

Equilibrium distribution functions is devised as function of motion integrals $f_{0\alpha}(\vec{P}) = f_{0\alpha}(W, P_y, P_x)$, where the total energy and generalized momentum take the form of:

$$W = \frac{1}{2}m_\alpha(v_x^2 + v_y^2 + v_z^2) + e_\alpha \phi(z),$$
$$P_y = m_\alpha v_y + e_\alpha A_y(z), \quad (4)$$
$$P_x = m_\alpha v_x.$$

Here, $\phi(z), A_y(z)$ - electrical and magnetic potentials $\vec{E}_0 = -grad\phi, \vec{B}_0 = rot\vec{A}$

At the first stage we study instability of a current sheath in electrically neutral approximation; therefore, as an equilibrium distribution function we will choose Harris:

$$f_{0\alpha}(W, P_y, P_x) = \left(\frac{m_\alpha}{2\pi\theta_\alpha}\right)^{\frac{3}{2}} n_0 \exp\{-\frac{m_\alpha}{2\theta_\alpha}[\alpha_1^2 + (\alpha_2 - V_\alpha)^2 + \alpha_3^2]\}. \quad (5)$$

Here, $V_i$ and $V_e$ are some average velocities of ions and electrons in Y-direction, and integrals of movement $\alpha_1, \alpha_2, \alpha_3$ are derived by rearrangement of expressions (4):

$$\alpha_1^2 = v_z^2 - \frac{2e_\alpha}{m_\alpha}v_y A_y - \frac{e_\alpha^2}{m_\alpha^2}A_y^2 + \frac{2e_\alpha}{m_\alpha}\phi,$$

$$\alpha_2 = v_y + \frac{e_\alpha}{m_\alpha}A_y,$$

$$\alpha_3 = v_x,$$

It is known that in a system of coordinates where the relation is fulfilled:

$$\frac{V_i}{\theta_i} = -\frac{V_e}{\theta_e}, \quad (6)$$

as solutions of current sheath for electromagnetic potentials and fields serve simple analytical

relations:

$$\begin{cases} \phi(z) = 0, \\ A_y(z) = -\dfrac{2\theta_i}{eV_i} \ln ch(\dfrac{z}{\lambda}), \end{cases} \quad (7)$$

$$B_0(z) = B_{0\max} th(\dfrac{z}{\lambda}), \quad (8)$$

$$\lambda = [\dfrac{2\theta_i}{\mu_0 e^2 n_0 V_i^2 (1+\theta_e/\theta_i)}]^{\frac{1}{2}} \text{ - characteristic thickness of Harris's layer.} \quad (9)$$

**Determination of non-equilibrium distribution function $\delta f_\alpha$**

Knowing now equilibrium distribution function (5), it is possible to derive equation for determination of non-equilibrium component to distribution function $\delta f_\alpha$:

$$\dfrac{\partial \delta f_\alpha}{\partial \varphi} - \dfrac{e_\alpha E_{0z} \sin\varphi}{\Omega_\alpha(z)} \dfrac{\partial \delta f_\alpha}{\partial P_{\alpha\perp}} = -\dfrac{i(\omega - \vec{k}\vec{v})}{\Omega_\alpha(z)} \delta f_\alpha + \dfrac{e_\alpha}{\Omega_\alpha(z)} \delta\vec{E} \dfrac{\partial f_{0\alpha}}{\partial \vec{P}_\alpha}. \quad (10)$$

Here, the following formula was used:

$E_{0\perp} = E_{0z} \sin\varphi$,

the Larmor frequency was entered:

$$\Omega_\alpha(z) = \dfrac{e_\alpha B_0(z)}{m_\alpha} \quad (11)$$

also Fourier expansion was used:

$$\delta f_\alpha(\vec{P},\vec{r},z,t) = \delta f_\alpha(\vec{P},\omega,\vec{k},z)\exp(-i\omega t + ik_x x + ik_y y),$$
$$\delta\vec{E}(\vec{r},z,t) = \delta\vec{E}(\omega,\vec{k},z)\exp(-i\omega t + ik_x x + ik_y y), \quad (12)$$
$$\delta\vec{B}(\vec{r},z,t) = \delta\vec{B}(\omega,\vec{k},z)\exp(-i\omega t + ik_x x + ik_y y).$$

Solution of equation (1) for the case of neglect of plasma polarization is known as [1]:

$$\delta f_\alpha = \dfrac{e_\alpha}{\Omega_\alpha(z)} \int_{\infty}^{\varphi} (\delta\vec{E} \dfrac{\partial f_{0\alpha}}{\partial \vec{P}_\alpha})_{\varphi'} \exp[\dfrac{i}{\Omega_\alpha(z)} \int_{\varphi}^{\varphi'} (\omega - \vec{k}\vec{v})_{\varphi''} d\varphi''] d\varphi'. \quad (13)$$

Theory of equilibrium current sheath is developed based on the equilibrium distribution function. Non-equilibrium distribution function (13) can be taken as a basis for analysis of boundary stability.

**Derivation of tensor of dielectric permeability and dispersion equation**

Inserting the determined non-equilibrium correction for distribution function $\delta f_\alpha$ (13) in the equation of current density induced in plasma:

$$\vec{j} = \sum_{\alpha} e_{\alpha} \int \vec{v} \delta f_{\alpha} d\vec{P},$$

and using the material equation of medium in the form of:

$$j_i = \sigma_{ij} \delta E_j, \tag{14}$$

it is possible to calculate the tensor of dielectric conductivity $\sigma_{ij}$, and then using formula:

$$\varepsilon_{ij} = \delta_{ij} + \frac{i}{\varepsilon_0 \omega} \sigma_{ij},$$

we may calculate tensor of dielectric permeability:

$$\varepsilon_{ij} = \delta_{ij} + \frac{i}{\varepsilon_0 \omega} \sigma_{ij} = \delta_{ij} + \frac{i}{\varepsilon_0 \omega} \sum_{\alpha} \frac{e_{\alpha}^2}{m_{\alpha} \Omega_{\alpha}(z)} \int d\vec{v} v_i \int_{\infty}^{\varphi} d\varphi' (\frac{\partial f_{0\alpha}}{\partial v_j})_{\varphi'} \exp[\frac{i}{\Omega_{\alpha}(z)} \int_{\varphi}^{\varphi'} d\varphi''(\omega - \vec{k}\vec{v})_{\varphi'}]. \tag{15}$$

Let's first integrate on angles. If we take the last integral, relation (15) will be converted as follows:

$$\varepsilon_{ij} = \delta_{ij} + \frac{i}{\varepsilon_0 \omega} \sum_{\alpha} \frac{e_{\alpha}^2}{m_{\alpha} \Omega_{\alpha}(z)} \int d\vec{v} \exp[-i\frac{\omega - k_x v_x}{\Omega_{\alpha}(z)} \varphi] \exp[ib_{\alpha} \sin\varphi] v_i \int_{\infty}^{\varphi} (\frac{\partial f_{0\alpha}}{\partial v_j})_{\varphi'} \cdot$$
$$\exp[i\frac{\omega - k_x v_x}{\Omega_{\alpha}(z)} \varphi'] \exp[-ib_{\alpha} \sin\varphi'] d\varphi', \tag{16}$$

where:

$$b_{\alpha} = \frac{k_{\perp} v_{\perp}}{\Omega_{\alpha}(z)}. \tag{17}$$

Let's describe in detail the calculation of one of the components of tensor of dielectric permeability, for example, component $\varepsilon_{xx}$.

Having inserted the derivative with velocity from equilibrium distribution function (5) in (16), and having used expansion:

$$\exp[-ib_{\alpha} \sin\varphi'] = \sum_{n=-\infty}^{\infty} J_n(b_{\alpha}) \exp[-in\varphi'], \tag{18}$$

where $J_n$ is Bessel function of n-order,

it is possible to obtain:

$$\varepsilon_{xx} = \delta_{xx} + A \exp[-\frac{m_{\alpha}}{2\theta_{\alpha}}(v_x^2 + v_{\perp}^2)] \exp(-\frac{m_{\alpha}^2 V_{\alpha}^2 - 2e_{\alpha} A_y V_{\alpha} m_{\alpha}}{2m_{\alpha} \theta_{\alpha}}) \cdot$$
$$\sum_n J_n(b_{\alpha}) \int_{\infty}^{\varphi} d\varphi' \exp[i(-n + \frac{\omega - k_x v_x}{\Omega_{\alpha}(z)})\varphi'] \exp(\lambda_{\alpha} \cos\varphi'). \tag{19}$$

Here,

$$A = \frac{i}{\varepsilon_0 \omega} \sum_{\alpha} \frac{e_{\alpha}^2}{\Omega_{\alpha}(z)} \left(\frac{m_{\alpha}}{2\pi\theta_{\alpha}}\right)^{\frac{3}{2}} n_{0\alpha} \int d\vec{v} \exp[-i\frac{\omega - k_x v_x}{\Omega_{\alpha}(z)} \varphi] \exp[ib_{\alpha} \sin\varphi](-\frac{v_x^2}{\theta_{\alpha}}), \tag{20}$$

$$\lambda_\alpha = \frac{m_\alpha V_\alpha v_\perp}{\theta_\alpha}. \tag{21}$$

By expansion of this function under integral (19) into Fourier series

$$\exp(\lambda_\alpha \cos\varphi') = \sum_{s=-\infty}^{\infty} C_s \exp(is\varphi'), \tag{22}$$

where:

$$C_s^{xx} = \frac{2}{\pi} \int_{\frac{\pi}{2}(\alpha-1)}^{\frac{\pi}{2}\alpha} \exp(-i4s\varphi') \exp[\lambda_\alpha \cos\varphi'] d\varphi' \tag{23}$$

let's derive from (19):

$$\varepsilon_{xx} = 1 + A \exp[-\frac{m_\alpha}{2\theta_\alpha}(v_x^2 + v_\perp^2)] \exp[-\frac{m_\alpha^2 V_\alpha^2 - 2e_\alpha A_y V_\alpha m_\alpha}{2m_\alpha \theta_\alpha}] \cdot$$

$$\sum_n J_n(b_\alpha) \sum_s C_s \frac{\exp[i(k-n+\frac{\omega-k_x v_x}{\Omega_\alpha(z)})\varphi]}{i(k-n+\frac{\omega-k_x v_x}{\Omega_\alpha(z)})}. \tag{24}$$

At integration it has been considered that lower limit of integration $\varphi = \infty$ value of the recovered function goes to zero as frequency possesses small imaginary positive part $\omega \to \omega + i\delta$ (the adiabatic hypothesis: perturbation of distribution function $\delta f_\alpha$ must disappear at $t \to -\infty$ in the accepted time dependency $\delta f_\alpha \approx \exp(-i\omega t)$ ; since disappearance means that frequency possesses at least infinitely small positive imaginary component).

We will perform speed integration in the system of cylindrical coordinates $d\vec{v} = dv_x v_\perp dv_\perp d\varphi$. Having inserted the value of A coefficient (20) and having performed the necessary transformations, we will derive from (24):

$$\varepsilon_{xx} = 1 - \sum_\alpha B_\alpha \sum_s \sum_l G_{xx}(s,l,m_\alpha,\theta_\alpha,k_\perp,\Omega_\alpha(z)) I_{xx}(\beta). \tag{25}$$

Here,

$$B_\alpha = \frac{2\pi e_\alpha^2}{\varepsilon_0 \omega} \frac{n_{0\alpha}}{(2\pi m_\alpha \theta_\alpha)^{\frac{3}{2}}} \frac{m_\alpha^3}{\theta_\alpha} \exp[-\frac{m_\alpha^2 V_\alpha^2 - 2e_\alpha m_\alpha V_\alpha A_y(z)}{2m_\alpha \theta_\alpha}], \tag{26}$$

$$C_s^{xx} = \frac{2}{\pi} \int_{\frac{\pi}{2}(\alpha-1)}^{\frac{\pi}{2}\alpha} \exp(-i4s\varphi') \exp[\frac{m_\alpha V_\alpha v_\perp}{\theta_\alpha} \cos\varphi'] d\varphi', \tag{27}$$

$$G_{xx}(s,l,m_\alpha,\theta_\alpha,k_\perp,\Omega_\alpha(z)) = \int_0^\infty C_s^{xx} v_\perp \exp(-\frac{m_\alpha v_\perp^2}{2\theta_\alpha}) J_{s+l}(\frac{k_\perp v_\perp}{\Omega_\alpha(z)}) J_l(\frac{k_\perp v_\perp}{\Omega_\alpha(z)}) dv_\perp, \qquad (28)$$

$$I_{xx}(\beta) = \int_{-\infty}^{+\infty} \frac{v_x^2 \exp(-\frac{m_\alpha v_x^2}{2\theta_\alpha})}{(\omega - k_x v_x - l\Omega_\alpha(z))} dv_x = \frac{\theta_\alpha}{m_\alpha k_x}[\frac{\sqrt{2\pi}}{\beta} - i\pi\beta^2 \exp(-\frac{\beta^2}{2})], \text{если } \beta \gg 1, \qquad (29)$$

where:

$$\beta = \frac{\omega - l\Omega_\alpha(z)}{k_x}\sqrt{\frac{m_\alpha}{\theta_\alpha}}. \qquad (30)$$

Here, coefficient $\alpha = i$ is for ions and $\alpha = e$ for electrons; $e_\alpha, m_\alpha, \theta_\alpha$ - charge, mass and temperature of the corresponding component of plasma; $A_y(z)$ - magnetic potential of the corresponding equilibrium solution; $k_x, k_y$ - components of wave vector of perturbations to be studied; $k_\perp = \sqrt{k_x^2 + k_y^2}$.

The remaining 8 components of tensor of dielectric permeability $\varepsilon_{ij}$ are calculated in a similar way. They are shown in Application.

The dispersion relation looks as follows [1]:

$$k^2 \delta_{ij} - k_i k_j - \frac{\omega^2}{c^2}\varepsilon_{ij}(\omega,\vec{k}) = 0. \qquad (31)$$

Components of tensor of dielectric permeability of sharply inhomogeneous plasma of the current sheath are determined by formula (25) and Application formulas.

**Conclusion**

The procedure of comprehensive analysis of instability of current sheathes in a wide range of frequencies and wave lengths in the electrically neutral approximation has been developed. This comprehensive analysis of instability is based on the consecutive solution of the kinetic equation with a self-consistent electromagnetic field (Vlasov's equation). For the first time the tensor of dielectric permeability is calculated and dispersion equation for sharply inhomogeneous plasma of current sheath is introduced.

# REFERENCIES

**Application**

$$\varepsilon_{xy} = -\sum_{\alpha} B_{\alpha} \sum_{s} \sum_{l} G_{xy}(s,l,m_{\alpha},\theta_{\alpha},k_{\perp},\Omega_{\alpha}(z)) I_{xy}(\beta). \tag{A1}$$

Here,

$$G_{xy}(s,l,m_{\alpha},\theta_{\alpha},k_{\perp},\Omega_{\alpha}(z)) = \int_{0}^{\infty} C_{s}^{xy} v_{\perp} \exp(-\frac{m_{\alpha}v_{\perp}^{2}}{2\theta_{\alpha}}) J_{s+l}(\frac{k_{\perp}v_{\perp}}{\Omega_{\alpha}(z)}) J_{l}(\frac{k_{\perp}v_{\perp}}{\Omega_{\alpha}(z)}) dv_{\perp},$$

$$I_{xy}(\beta) = \int_{-\infty}^{+\infty} \frac{v_{x} \exp(-\frac{m_{\alpha}v_{x}^{2}}{2\theta_{\alpha}})}{(\omega - k_{x}v_{x} - l\Omega_{\alpha}(z))} dv_{x} \approx \frac{1}{k_{x}} \sqrt{\frac{\theta_{\alpha}}{m_{\alpha}}} \frac{\sqrt{2\pi}}{\beta^{2}}, \; \beta \gg 1,$$

$$C_{s}^{xy} = \frac{2}{\pi} \int_{\frac{\pi}{2}(\alpha-1)}^{\frac{\pi}{2}\alpha} \exp(-i4s\varphi') \cdot (v_{\perp}\cos\varphi' - V_{\alpha}) \cdot \exp(\frac{m_{\alpha}V_{\alpha}v_{\perp}}{2\theta_{\alpha}}\cos\varphi') d\varphi'.$$

$$\varepsilon_{xz} = -\sum_{\alpha} B_{\alpha} \sum_{s} \sum_{l} G_{xz}(s,l,m_{\alpha},\theta_{\alpha},k_{\perp},\Omega_{\alpha}(z)) I_{xz}(\beta). \tag{A2}$$

Here,

$$G_{xz}(s,l,m_{\alpha},\theta_{\alpha},k_{\perp},\Omega_{\alpha}(z)) = \int_{0}^{\infty} C_{s}^{xz} v_{\perp}^{2} \exp(-\frac{m_{\alpha}v_{\perp}^{2}}{2\theta_{\alpha}}) J_{s+l}(\frac{k_{\perp}v_{\perp}}{\Omega_{\alpha}(z)}) J_{l}(\frac{k_{\perp}v_{\perp}}{\Omega_{\alpha}(z)}) dv_{\perp},$$

$$I_{xz}(\beta) = \int_{-\infty}^{+\infty} \frac{v_{x} \exp(-\frac{m_{\alpha}v_{x}^{2}}{2\theta_{\alpha}})}{(\omega - k_{x}v_{x} - l\Omega_{\alpha}(z))} dv_{x} \approx \frac{1}{k_{x}} \sqrt{\frac{\theta_{\alpha}}{m_{\alpha}}} \frac{\sqrt{2\pi}}{\beta^{2}}, \; \beta \gg 1,$$

$$C_{s}^{xz} = \frac{2}{\pi} \int_{\frac{\pi}{2}(\alpha-1)}^{\frac{\pi}{2}\alpha} \exp(-i4s\varphi') \cdot \sin\varphi' \cdot \exp(\frac{m_{\alpha}V_{\alpha}v_{\perp}}{2\theta_{\alpha}}\cos\varphi') d\varphi'.$$

$$\varepsilon_{yx} = -\sum_{\alpha} B_{\alpha} \sum_{s} \sum_{l} G_{yx}(s,l,m_{\alpha},\theta_{\alpha},k_{\perp},\Omega_{\alpha}(z)) I_{yx}(\beta). \tag{A3}$$

Here,

$$G_{yx}(s,l,m_{\alpha},\theta_{\alpha},k_{\perp},\Omega_{\alpha}(z)) = \int_{0}^{\infty} C_{s}^{yx} v_{\perp}^{2} \exp(-\frac{m_{\alpha}v_{\perp}^{2}}{2\theta_{\alpha}}) J_{s+l}(\frac{k_{\perp}v_{\perp}}{\Omega_{\alpha}(z)}) l J_{l}(\frac{k_{\perp}v_{\perp}}{\Omega_{\alpha}(z)}) \frac{\Omega_{\alpha}(z)}{k_{\perp}v_{\perp}} dv_{\perp},$$

$$I_{yx}(\beta) = \int_{-\infty}^{+\infty} \frac{v_{x} \exp(-\frac{m_{\alpha}v_{x}^{2}}{2\theta_{\alpha}})}{(\omega - k_{x}v_{x} - l\Omega_{\alpha}(z))} dv_{x} \approx \frac{1}{k_{x}} \sqrt{\frac{\theta_{\alpha}}{m_{\alpha}}} \frac{\sqrt{2\pi}}{\beta^{2}}, \; \beta \gg 1,$$

$$C_s^{yx} = \frac{2}{\pi} \int_{\frac{\pi}{2}(\alpha-1)}^{\frac{\pi}{2}\alpha} \exp(-i4s\varphi') \cdot \exp(\frac{m_\alpha V_\alpha v_\perp}{2\theta_\alpha}\cos\varphi')d\varphi'.$$

$$\varepsilon_{yy} = 1 - \sum_\alpha B_\alpha \sum_s \sum_l G_{yy}(s,l,m_\alpha,\theta_\alpha,k_\perp,\Omega_\alpha(z))I_{yy}(\beta). \quad (A4)$$

Here,

$$G_{yy}(s,l,m_\alpha,\theta_\alpha,k_\perp,\Omega_\alpha(z)) = \int_0^\infty C_s^{yy} v_\perp^2 \exp(-\frac{m_\alpha v_\perp^2}{2\theta_\alpha}) J_{s+l}(\frac{k_\perp v_\perp}{\Omega_\alpha(z)}) lJ_l(\frac{k_\perp v_\perp}{\Omega_\alpha(z)}) \frac{\Omega_\alpha(z)}{k_\perp v_\perp} dv_\perp,$$

$$I_{yy}(\beta) = \int_{-\infty}^{+\infty} \frac{\exp(-\frac{m_\alpha v_x^2}{2\theta_\alpha})}{(\omega - k_x v_x - l\Omega_\alpha(z))} dv_x \approx \frac{1}{k_x}\frac{\sqrt{2\pi}}{\beta}, \quad \beta >> 1,$$

$$C_s^{yy} = \frac{2}{\pi} \int_{\frac{\pi}{2}(\alpha-1)}^{\frac{\pi}{2}\alpha} \exp(-i4s\varphi') \cdot (v_\perp \cos\varphi' - V_\alpha) \cdot \exp(\frac{m_\alpha V_\alpha v_\perp}{2\theta_\alpha}\cos\varphi')d\varphi'.$$

$$\varepsilon_{yz} = -\sum_\alpha B_\alpha \sum_s \sum_l G_{yz}(s,l,m_\alpha,\theta_\alpha,k_\perp,\Omega_\alpha(z)) \cdot I_{yz}(\beta). \quad (A5)$$

Here,

$$G_{yz}(s,l,m_\alpha,\theta_\alpha,k_\perp,\Omega_\alpha(z)) = \int_0^\infty C_s^{yz} v_\perp^3 \exp(-\frac{m_\alpha v_\perp^2}{2\theta_\alpha}) J_{s+l}(\frac{k_\perp v_\perp}{\Omega_\alpha(z)}) lJ_l(\frac{k_\perp v_\perp}{\Omega_\alpha(z)}) \frac{\Omega_\alpha(z)}{k_\perp v_\perp} dv_\perp,$$

$$I_{yz}(\beta) = \int_{-\infty}^{+\infty} \frac{\exp(-\frac{m_\alpha v_x^2}{2\theta_\alpha})}{(\omega - k_x v_x - l\Omega_\alpha(z))} dv_x \approx \frac{1}{k_x}\frac{\sqrt{2\pi}}{\beta}, \quad \beta >> 1,$$

$$C_s^{yz} = \frac{2}{\pi} \int_{\frac{\pi}{2}(\alpha-1)}^{\frac{\pi}{2}\alpha} \exp(-i4s\varphi') \cdot \sin\varphi' \cdot \exp(\frac{m_\alpha V_\alpha v_\perp}{2\theta_\alpha}\cos\varphi')d\varphi'.$$

$$\varepsilon_{zx} = i\sum_\alpha B_\alpha \sum_s \sum_l G_{zx}(s,l,m_\alpha,\theta_\alpha,k_\perp,\Omega_\alpha(z)) \cdot I_{zx}(\beta). \quad (A6)$$

Here

$$G_{zx}(s,l,m_\alpha,\theta_\alpha,k_\perp,\Omega_\alpha(z)) = \int_0^\infty C_s^{zx} v_\perp^2 \exp(-\frac{m_\alpha v_\perp^2}{2\theta_\alpha}) J_{s+l}(\frac{k_\perp v_\perp}{\Omega_\alpha(z)}) J_l'(\frac{k_\perp v_\perp}{\Omega_\alpha(z)}) dv_\perp$$

$$J_l'(\frac{k_\perp v_\perp}{\Omega_\alpha(z)}) = \frac{1}{2}(J_{l-1}(\frac{k_\perp v_\perp}{\Omega_\alpha(z)}) - J_{l+1}(\frac{k_\perp v_\perp}{\Omega_\alpha(z)})).$$

$$I_{zx}(\beta) = \int_{-\infty}^{+\infty} \frac{v_x \exp(-\frac{m_\alpha v_x^2}{2\theta_\alpha})}{(\omega - k_x v_x - l\Omega_\alpha(z))} dv_x \approx \frac{1}{k_x}\sqrt{\frac{\theta_\alpha}{m_\alpha}}\frac{\sqrt{2\pi}}{\beta^2}, \quad \beta \gg 1,$$

$$C_s^{zx} = \frac{2}{\pi}\int_{\frac{\pi}{2}(\alpha-1)}^{\frac{\pi}{2}\alpha} \exp(-i4s\varphi')\exp(\frac{m_\alpha V_\alpha v_\perp}{2\theta_\alpha}\cos\varphi')d\varphi'.$$

$$\varepsilon_{zy} = i\sum_\alpha B_\alpha \sum_s \sum_l G_{zy}(s,l,m_\alpha,\theta_\alpha,k_\perp,\Omega_\alpha(z)) \cdot I_{zy}(\beta). \tag{A7}$$

Here,

$$G_{zy}(s,l,m_\alpha,\theta_\alpha,k_\perp,\Omega_\alpha(z)) = \int_0^\infty C_s^{zy} v_\perp^2 \exp(-\frac{m_\alpha v_\perp^2}{2\theta_\alpha})J_{s+l}(\frac{k_\perp v_\perp}{\Omega_\alpha(z)})J'_l(\frac{k_\perp v_\perp}{\Omega_\alpha(z)})dv_\perp,$$

Where:

$$J'_l(\frac{k_\perp v_\perp}{\Omega_\alpha(z)}) = \frac{1}{2}(J_{l-1}(\frac{k_\perp v_\perp}{\Omega_\alpha(z)}) - J_{l+1}(\frac{k_\perp v_\perp}{\Omega_\alpha(z)})).$$

$$I_{zy}(\beta) = \int_{-\infty}^{+\infty} \frac{\exp(-\frac{m_\alpha v_x^2}{2\theta_\alpha})}{(\omega - k_x v_x - l\Omega_\alpha(z))} dv_x \approx \frac{1}{k_x}\frac{\sqrt{2\pi}}{\beta}, \quad \beta \gg 1,$$

$$C_s^{zy} = \frac{2}{\pi}\int_{\frac{\pi}{2}(\alpha-1)}^{\frac{\pi}{2}\alpha} \exp(-i4s\varphi') \cdot (v_\perp \cos\varphi' - V_\alpha) \cdot \exp(\frac{m_\alpha V_\alpha v_\perp}{2\theta_\alpha}\cos\varphi')d\varphi'.$$

$$\varepsilon_{zz} = 1 + i\sum_\alpha B_\alpha \sum_s \sum_l G_{zz}(s,l,m_\alpha,\theta_\alpha,k_\perp,\Omega_\alpha(z)) \cdot I_{zz}(\beta). \tag{A8}$$

Here,

$$G_{zz}(s,l,m_\alpha,\theta_\alpha,k_\perp,\Omega_\alpha(z)) = \int_0^\infty C_s^{zz} v_\perp^3 \exp(-\frac{m_\alpha v_\perp^2}{2\theta_\alpha})J_{s+l}(\frac{k_\perp v_\perp}{\Omega_\alpha(z)})J'_l(\frac{k_\perp v_\perp}{\Omega_\alpha(z)})dv_\perp,$$

Where:

$$J'_l(\frac{k_\perp v_\perp}{\Omega_\alpha(z)}) = \frac{1}{2}(J_{l-1}(\frac{k_\perp v_\perp}{\Omega_\alpha(z)}) - J_{l+1}(\frac{k_\perp v_\perp}{\Omega_\alpha(z)})).$$

$$I_{zz}(\beta) = \int_{-\infty}^{+\infty} \frac{\exp(-\frac{m_\alpha v_x^2}{2\theta_\alpha})}{(\omega - k_x v_x - l\Omega_\alpha(z))} dv_x \approx \frac{1}{k_x}\frac{\sqrt{2\pi}}{\beta}, \quad \beta \gg 1,$$

$$C_s^{zz} = \frac{2}{\pi}\int_{\frac{\pi}{2}(\alpha-1)}^{\frac{\pi}{2}\alpha} \exp(-i4s\varphi') \cdot \sin\varphi' \cdot \exp(\frac{m_\alpha V_\alpha v_\perp}{2\theta_\alpha}\cos\varphi')d\varphi'.$$